\documentclass[prb,twocolumn,amsmath,amssymb]{revtex4}
\usepackage{natbib}
\usepackage{graphicx}         
\usepackage{dcolumn}         
\usepackage{bm}                 

\begin{document}

\title{Enhanced spin-orbit scattering length in narrow Al$_x$Ga$_{1-x}$N/GaN wires}

\author{Patrick Lehnen}
\affiliation{Institute of Bio- and Nanosystems (IBN-1),cni-Center of
Nanoelectronic Systems for Information Technology, and Virtual
Institute of Spinelectronics (VISel), Research Centre J\"ulich GmbH,
52425 J\"ulich, Germany}

\author{Thomas Sch\"apers}
\email{th.schaepers@fz-juelich.de} \affiliation{Institute of Bio-
and Nanosystems (IBN-1),cni-Center of Nanoelectronic Systems for
Information Technology, and Virtual Institute of Spinelectronics
(VISel), Research Centre J\"ulich GmbH, 52425 J\"ulich, Germany}

\author{Nicoleta Kaluza}
\affiliation{Institute of Bio- and Nanosystems (IBN-1),cni-Center of
Nanoelectronic Systems for Information Technology, and Virtual
Institute of Spinelectronics (VISel), Research Centre J\"ulich GmbH,
52425 J\"ulich, Germany}

\author{Nicolas Thillosen}
\affiliation{Institute of Bio- and Nanosystems (IBN-1),cni-Center of
Nanoelectronic Systems for Information Technology, and Virtual
Institute of Spinelectronics (VISel), Research Centre J\"ulich GmbH,
52425 J\"ulich, Germany}

\author{Hilde Hardtdegen}
\affiliation{Institute of Bio- and Nanosystems (IBN-1),cni-Center of
Nanoelectronic Systems for Information Technology, and Virtual
Institute of Spinelectronics (VISel), Research Centre J\"ulich GmbH,
52425 J\"ulich, Germany}

\date{\today}

\hyphenation{GaInAs}

\begin{abstract}
The magnetotransport in a set of identical parallel
Al$_x$Ga$_{1-x}$N/GaN quantum wire structures was investigated.
The width of the wires was ranging between 1110~nm and 340~nm. For
all sets of wires clear Shubnikov--de Haas oscillations are
observed. We find that the electron concentration and mobility is
approximately the same for all wires, confirming that the electron
gas in the Al$_x$Ga$_{1-x}$N/GaN heterostructure is not
deteriorated by the fabrication procedure of the wire structures.
For the wider quantum wires the weak antilocalization effect is
clearly observed, indicating the presence of spin-orbit coupling.
For narrow quantum wires with an effective electrical width below
250~nm the weak antilocalization effect is suppressed. By
comparing the experimental data to a theoretical model for quasi
one-dimensional structures we come to the conclusion that the
spin-orbit scattering length is enhanced in narrow wires.
\end{abstract}

\maketitle

\section{Introduction}

The Al$_x$Ga$_{1-x}$N/GaN material system is a very promising
candidate for future spin electronic applications. The reason is,
that two important requirements for the realization of spin
electronic devices are fulfilled in this material class. First,
transition-metal-doped GaN diluted magnetic semiconductors have been
shown to have high Curie temperatures for injection and detection of
spin polarized carriers (see, e.g. Ref.~\onlinecite{Liu05} and
references therein) and second, spin-orbit coupling for spin control
in non-magnetic Al$_x$Ga$_{1-x}$N/GaN heterostructures was
observed.\cite{Lo02,Tsubaki02,Lu04,Cho05,Weber05,Thillosen06,Schmult06a,Thillosen06a,Schmult06,Kurdak06}

Spin-orbit coupling in Al$_x$Ga$_{1-x}$N/GaN two-dimensional
electron gases (2DEGs) can be investigated by analyzing the
characteristic beating pattern in Shubnikov--de Haas
oscillations,\cite{Lo02,Tsubaki02,Lu04,Cho05} by measuring the
circular photogalvanic effect,\cite{Weber05} or by studying
weak-antilocalization.\cite{Lu04,Thillosen06,Thillosen06a,Schmult06,Kurdak06}
The latter is an electron interference effect where the random
deviations of the spin orientations between time reversed paths
result in an enhanced
conductance.\cite{Hikami80,Bergmann82b,Gusev84} From weak
antilocalization measurements information on characteristic length
scales, i.e. the spin-orbit scattering length $l_{so}$ and the phase
coherence length $l_\phi$, can be obtained.

For quasi one-dimensional systems it was predicted
theoretically\cite{Bournel98,Malshukov00,Kiselev00,Pareek02} and
shown
experimentally\cite{Schaepers06,Wirthmann06,Holleitner06,Kwon07}
that $l_{so}$ can be considerably enhanced compared to the value
of the 2DEG. This has important implications for the performance
of spin electronic devices, e.g. the spin
field-effect-transistor,\cite{Datta90} since an enhanced value of
$l_{so}$ results in a larger degree of spin polarization in the
channel and thus to larger signal
modulation.\cite{Datta90,Bournel98} In addition, many of the
recently proposed novel spin electronic device structures
explicitly make use of one-dimensional channels, because the
restriction to only one dimension allows new switching
schemes.\cite{Nitta99,Kiselev01,Governale02b,Cummings06}

Very recently, transport measurements on AlGaN/GaN-based
one-dimensional structures, i.e quantum points contacts, have been
reported.\cite{Chou05,Schmult06a} With respect to possible spin
electronic applications it is of great interest, how the spin
transport takes place in AlGaN/GaN quasi one-dimensional structures.
Since an enhanced value of $l_{so}$ is very advantageous for the
design of spin electronic devices, it would be very desirable if
this effect can be observed in Al$_x$Ga$_{1-x}$N/GaN wire
structures.

Here, we report on magnetotransport measurements on
Al$_x$Ga$_{1-x}$N/GaN parallel quantum wire structures. We will
begin by discussing the basic transport properties of wires with
different widths, i.e. resistivity, sheet electron concentration,
and mobility. Spin-orbit coupling in our Al$_x$Ga$_{1-x}$N/GaN
quantum wires is investigated by analyzing the weak antilocalization
effect. We will discuss to which extent the weak antilocalization
effect in Al$_x$Ga$_{1-x}$N/GaN heterostructures is affected by the
additional confinement in wire structures. By fitting a theoretical
model to or experimental data, we will be able to answer the
question if the spin-orbit scattering length increases with
decreasing wire width, as found in quantum wires fabricated from
other types of heterostructures.

\section{Experimental}

\begin{figure}
\includegraphics[width=\columnwidth]{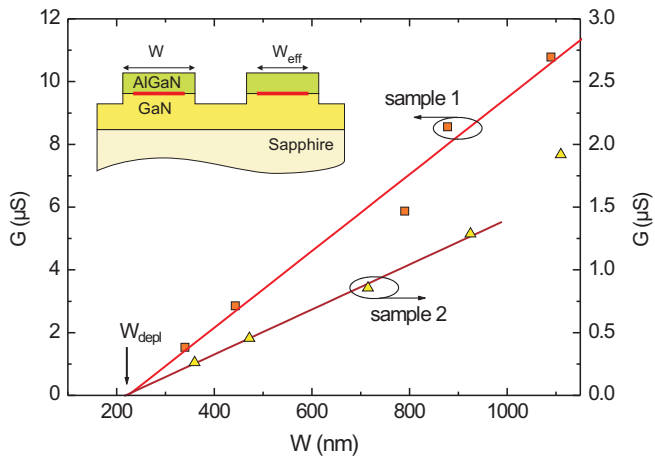}
\caption{(Color online) Conductance $G$ as a function of the
geometrical width $W$. The conductance of a single wire is
plotted, which was determined by dividing the total conductance by
the number of wires connected in parallel. The full lines
represent the corresponding linear fits. The arrows indicates the
total width of the depletion zones $W_{depl}$. The inset shows a
schematics of the cross-section of the wires. Here, $W$
corresponds to the geometrical width, while $W_{eff}$ indicates
the effective electrical width. \label{Fig1}}
\end{figure}

The AlGaN/GaN heterostructures were grown by metalorganic vapor
phase epitaxy on a (0001) Al$_2$O$_3$ substrate. Two different
samples were investigated. Sample~1 consisted of a 3-$\mu$m-thick
GaN layer followed by a 35-nm-thick Al$_{0.20}$Ga$_{0.80}$N top
layer, while in sample~2 a 40-nm-thick Al$_{0.10}$Ga$_{0.90}$N
layer was used as a top layer. The quantum wire structures were
prepared by first defining a Ti etching mask using electron beam
lithography and lift-off. Subsequently, the AlGaN/GaN wires were
formed by Ar$^+$ ion beam etching. The etching depth of 95~nm was
well below the depth of the AlGaN/GaN interface. The electron beam
lithography pattern was chosen so that a number of 160 identical
wires, each 620~$\mu$m long, were connected in parallel. A
schematic cross section of the parallel wires is shown in
Fig.~\ref{Fig1} (inset).

Different sets of wires were prepared comprising a geometrical width
$W$ ranging from 1110~nm down to 340~nm (see Table~\ref{Table1}).
The geometrical widths of the wires were determined by means of
scanning electron microscopy. The sample geometry with quantum wires
connected in parallel was chosen, in order to suppress universal
conductance fluctuations.\cite{Beenakker97} After removing the Ti
mask by HF, Ti/Al/Ni/Au Ohmic contacts were defined by optical
lithography. The Ohmic contacts were alloyed at 900$^\circ$C for
30~s. For reference purposes a 100-$\mu$m-wide Hall bar structure
with voltage probes separated by a distance of 410~$\mu$m were
prepared on the same chip.

The measurements were performed in a He-3 cryostat at temperatures
ranging from 0.4~K to 4.0~K. The resistances were measured by
employing a current-driven lock-in technique with an ac excitation
current of 100~nA and 1~$\mu$A for sample~1 and 2, respectively.

\begin{table}
\caption{Summary of characteristic parameters of both samples: The
sample number, geometrical wire width $W$, effective electrical wire
width $W_{eff}$, resisistivity $\rho$, sheet electron concentration
$n_{2D}$, mobility $\mu$, and elastic mean free path $l_{el}$. The
spin-orbit scattering length $l_{so}$, and phase coherence length
$l_\phi$ were extracted from the fit using the Kettemann
model.\cite{Kettemann07}
 \label{Table1}}
 \begin{ruledtabular}
 \begin{tabular}{cccccccccc}
 $\#$ & $W$ & $W_{eff}$ & $\rho $ & $n_{2D}$ & $\mu$ &$l_{el}$ & $l_{so}$ &  $l_\phi$ \\
 &(nm) & (nm) & ($\Omega$) &  ($10^{12}$cm$^{-2})$ & (cm$^2$/Vs) & (nm) &  (nm)  & (nm)
 \\ \colrule
1& 1090& 880& 131 & 5.1 & 9400 & 349    & 550   & 3000 \\
1& 880&  670& 126 & 5.2 & 9600 & 360    & 600   & 2950 \\
1& 690&  480& 132 & 4.9 & 9700 & 344    & 700   & 2500 \\
1& 440&  230& 132 & 5.2 & 9000 & 341    & 1300  & 1550 \\
1& 340&  130& 136 & 4.5 & 10000 & 343   & $>$1800 & 1150 \\ \\
2& 1110& 870& 730 & 2.2 & 4000 & 96 & 500   & 1200 \\
2& 930&  690& 860 & 2.2 & 3400 & 82 & 520   & 1000 \\
2& 720&  480& 900 & 2.0 & 3400 & 81 & 640   & 950 \\
2& 470&  230& 830 & 2.0 & 3800 & 88 & $>$850  & 900 \\
2& 360&  120& 740 & 1.9 & 4300 & 100 & $>$1000  & 670 \\
 \end{tabular}
 \end{ruledtabular}
 \end{table}

\section{Results and Discussion}

In order to gain information on the transport properties of the
Al$_x$Ga$_{1-x}$N/GaN layer systems, Shubnikov--de Haas
oscillations were measured on the Hall bar samples. At a
temperature of 0.5~K sheet electron concentrations $n_{2D}$ of
$5.1 \times 10^{12}$~cm$^{-2}$ and $2.2 \times 10^{12}$~cm$^{-2}$,
were determined for sample 1 and 2, respectively. The Fermi
energies calculated from $n_{2D}$ are 55~meV for sample 1 and
24~meV for sample 2. Here, an effective electron mass of
$m^*=0.22\; m_{e}$ was taken into account.\cite{Thillosen06a} The
mobilities $\mu$ were 9150~cm$^2$/Vs and 3930~cm$^2$/Vs for sample
1 and 2, respectively, resulting in elastic mean free paths
$l_{el}$ of 314~nm and 95~nm. The smaller electron concentration
of sample~2 can be attributed to the lower Al-content of the
Al$_x$Ga$_{1-x}$N barrier layer resulting in a smaller
polarization-doping.\cite{Ambacher00} The lower mobility found in
sample~2 compared to sample~1 can be explained by the reduced
screening at lower electron concentrations.\cite{Sakowicz06}

Owing to the large surface potential of GaN, which has been
determined to be between 0.5 and 0.6~eV,\cite{Kocan02} a
considerable surface carrier depletion can be expected. For our
wires the carrier depletion at the mesa edges will result in an
effective electrical width $W_{eff}$ which is smaller than the
measured geometrical width $W$. In order to gain information on the
lateral width of the depletion zone, the wire conductance at zero
magnetic field was determined for different wire widths. In
Fig.~\ref{Fig1} the single-wire conductance $G$ is shown as a
function of the wire width for both samples. It can be seen that for
both samples $G$ scales linearly with $W$. The total width of the
depletion zone was determined from the linear extrapolation to
$G=0$, indicated by $W_{depl}$ in
Fig.~\ref{Fig1}.\cite{Menschig90,Long93} The depletion zone width
for sample~1 is 210~nm while for sample~2 a value of 240~nm was
determined. The larger value of $W_{depl}$ for sample 2 can be
attributed to the lower electron concentration compared to sample~1.
The corresponding effective electrical width $W_{eff}$, defined by
$W-W_{depl}$, is listed in Table~\ref{Table1}. The two-dimensional
resistivity $\rho$ of the wires at $B=0$ was calculated based on
$W_{eff}$. As can be seen by the values of $\rho$ given in
Table~\ref{Table1}, for sample~1 the resistivity remains at
approximately the same value if the wire width is reduced. A similar
behavior is observed for sample~2, although the variations are
somewhat larger. In any case no systematic change of $\rho$ is found
for both samples.

\begin{figure}
\includegraphics[width=\columnwidth]{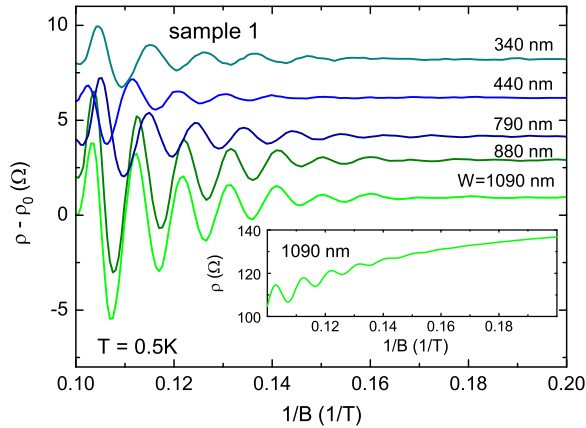}
\caption{(Color online) Magnetoresisitivity as a function of the
inverse magnetic field for set of wires of different widths
(sample~1). The slowly varying background resistivity $\rho_0(B)$
was subtracted. For clarity, the curves are offset by 2~$\Omega$.
The resistance of the sets of wires was measured at a temperature
of 0.5~K. The inset shows the resistivity of the 1090~nm~wide
wires before the background resistivity $\rho_0(B)$ was
subtracted.\label{Fig2}}
\end{figure}

As can be seen in Fig~\ref{Fig2}, clear Shubnikov--de Haas
oscillations in the magnetoresistivity $\rho(B)-\rho_0(B)$ are
resolved for different sets of wires of sample 1. For a better
comparison the slowly varying field-dependent background resistivity
$\rho_0(B)$ was subtracted. In order to get an impression on the
relation between the amplitude of the Shubnikov--de Haas
oscillations and the background resistivity, the total resistivity
$\rho(B)$ is shown exemplarily for the 1090-nm-wide wires in
Fig.~\ref{Fig2} (inset). As can be seen here, the oscillation
amplitude turns out to be small compared to $\rho_0(B)$, because of
the relatively low mobility. From the oscillation period of
$\rho(B)-\rho_0(B)$ vs. $1/B$ the sheet electron concentration
$n_{2D}$ was determined for the different sets of wires. As can be
seen in Fig.~\ref{Fig2}, the oscillation period and thus $n_{2D}$ is
approximately the same for all sets of wires (cf.
Table~\ref{Table1}). The values of $n_{2D}$ are comparable to the
value found for the 2DEG. As given in Table~\ref{Table1}, for
sample~2 the values of $n_{2D}$ for the different sets of wires were
also found to be close to the value extracted from the corresponding
Hall bar structure.

The mobility $\mu$ and elastic mean free path $l_{el}$ was
determined from $n_{2D}$ and $\rho(B=0)$. As can be inferred from
the values of $\mu$ and $l_{el}$ given in Table~\ref{Table1}, both
quantities are similar for all sets of wires for a given
heterostructure. For sample~2, $l_{el}$ is always smaller than
$W_{eff}$, therefore no significant deviation from the 2DEG
conductivity is expected. However, for the 440~nm and 340~nm wide
wires of sample~1, $l_{el}$ exceeds $W_{eff}$ so that a boundary
scattering contribution is expected. However, since the mobility
is not decreased, we can conclude that the boundary scattering is
predominately specular. Probably, the smooth potential from the
depletion zone favors specular reflection.

\begin{figure}
\includegraphics[width=0.7\columnwidth]{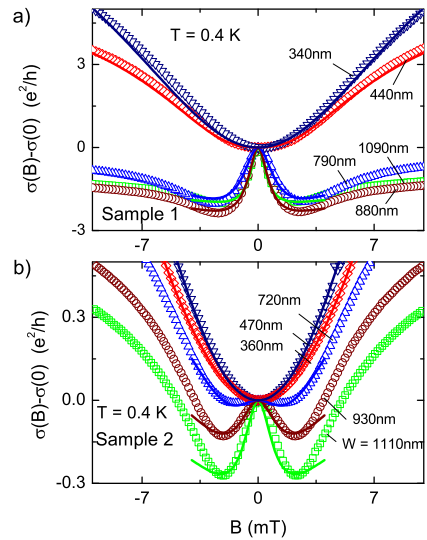}
\caption{(Color online) (a) Experimental magnetoconductivity
$\sigma(B)-\sigma(0)$ normalized to $e^2/h$ for different sets of
wires of sample~1. The measurement temperature was 0.4~K. Sets of
wires with a geometrical width ranging from 1090~nm down to 340~nm
were measured. The full lines show the calculated values using the
Kettemann model.\cite{Kettemann07} (b) Corresponding measurements
of $\sigma(B)-\sigma(0)$ for sets of wires of sample~2 with widths
in the range from 1110~nm to 360~nm. The full lines show the
calculated magnetoconductivity.\label{Fig3}}
\end{figure}

We now turn to the investigation of spin-related effects in the
electron transport. In Fig.~\ref{Fig3}(a) the normalized
magnetoconductivity $\sigma(B)-\sigma(0)$ is shown for different
sets of wires of sample~1. For the narrow wires with a width up to
440~nm the magnetoconductivity monotonously increases for increasing
values of $|B|$, which can be attributed to weak localization. The
weak localization effect originates from the constructive
interference of time-reversed pathes for the case when spin-orbit
scattering can be neglected. In contrast, for the 1090~nm, 880~nm,
and 790~nm wide wires, a peak is found in the magnetoconductivity at
$B=0$, which is due to weak antilocalization. The slope of the
magnetoconductivity changes sign at $|B|\approx~2.2$~mT. This value
corresponds well to the positions of the minima found in the weak
antilocalization measurements on the Hall bars of sample~1. For
magnetic fields beyond 2.2~mT the transport is governed by weak
localization, where the magnetoconductivity increases with $|B|$.

As can be seen in Fig.~\ref{Fig3}(b), a similar behavior is found
for sample~2. For wire widths up to 470~nm weak localization is
observed, whereas for the 1110~nm, 930~nm and 720~nm wide wires weak
antilocalization is found. In contrast to sample~1, the width of the
weak antilocalization peak depends on the widths of the wires. For
the first two sets of wires minima in $\sigma(B)-\sigma(0)$ are
found at $B = \pm2.2$~mT. Whereas, for the 720-nm-wide wires minima
are observed at $\pm1.5$~mT. The peak height due to weak
antilocalization decreases with decreasing wire width. In general,
the modulations of $\sigma(B)-\sigma(0)$ are found to be
considerably smaller for sample 2 compared to sample 1, which can be
attributed to the smaller elastic mean free path and, as it will be
shown later, to the smaller phase coherence length.

\begin{figure}[h]
\includegraphics[width=\columnwidth]{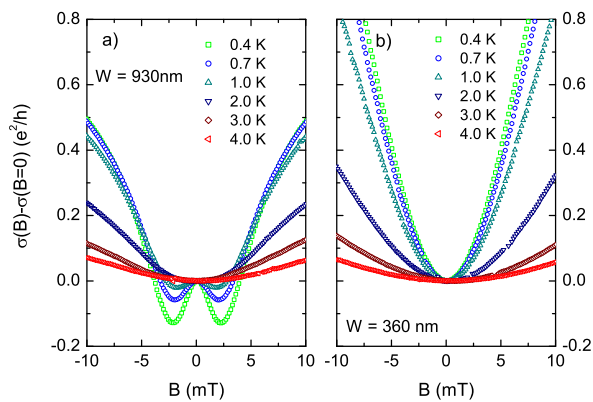}
\caption{(Color online) (a) Magnetoconductivity
$\sigma(B)-\sigma(0)$ normalized to $e^2/h$ of the 930-nm-wide set
of wires of sample 2 at different temperatures in the range from
0.4~K to 4~K. (b) Corresponding measurements for the set of
360-nm-wide wires. \label{Fig4}}
\end{figure}

With increasing temperature the weak antilocalization peak
decreases. This can be seen in Fig.~\ref{Fig4}(a), where
$\sigma(B)-\sigma(0)$ is shown at different temperatures for the
930-nm-wide wires of sample~2. Above 2~K no signature of weak
antilocalization is found anymore. Furthermore, the weak
localization contribution to $\sigma(B)-\sigma(0)$ successively
decreases with increasing temperature. This effect can be attributed
to the decreasing phase coherence length with increasing
temperature.\cite{Altshuler82,Choi87} As can be seen in
Fig.~\ref{Fig4}(b), for the 360-nm-wide wires only weak localization
was observed. Similar to the wider sets of wires, the weak
localization effect is damped with increasing temperatures.

From weak antilocalization measurements the characteristic length
scales, i.e. $l_\phi$ and $l_{so}$, can be estimated. In order to
get some reference value for the 2DEG, the model developed by
Iordanskii, Lyanda-Geller, and Pikus\cite{Iordanskii94} (ILP-model)
was fitted to the weak antilocalization measurements of the Hall bar
structures. Only the Rashba contribution was considered, here. For
sample~1, $l_\phi$ and $l_{so}$ were found to be 1980~nm and 300~nm
at 0.5~K, respectively, whereas for sample~2 the corresponding
values were 1220~nm and 295~nm at 0.4~K. For both samples the
effective spin-orbit coupling parameter $\alpha=\hbar^2/2m^*l_{so}$
is approximately $5.8 \times 10^{-13}$~eVm. The zero-field
spin-splitting energy can be estimated by using the the expression
$\Delta_{so}=2k_F \alpha$, with $k_F$ the Fermi wavenumber given by
$\sqrt{2 \pi n_{2D}}$. For sample~1 one obtains a value of
$\Delta_{so}=0.66$~meV, while for sample~2 one finds 0.43~meV. The
values of $\Delta_{so}$ are relatively large compared to their
corresponding Fermi energies, which confirms the presence of a
pronounced spin-orbit coupling in Al$_x$Ga$_{1-x}$N/GaN
2DEGs.\cite{Thillosen06,Thillosen06a,Schmult06,Kurdak06}

The ILP-model is only valid for 2DEGs with $l_\phi \ll W$, thus it
cannot be applied to our wire structures. Very recently, a model
appropriate for wire structures was developed by
Kettemann,\cite{Kettemann07} which covers the case $W < l_\phi$.
Here, the quantum correction to the conductivity is given by:
\begin{eqnarray}
\sigma(B)-\sigma(0) &=& \frac{e^2}{h}  \left(
\frac{\sqrt{H_W}}{\sqrt{H_\phi + B^*/4}} -
\frac{\sqrt{H_W}}{\sqrt{H_\phi + B^*/4 + H_{so}}} \right. \nonumber \\
&-& \left. 2 \frac{\sqrt{H_W}}{\sqrt{H_\phi + B^*/4 + H_{so}/2}}
\right) \; , \label{Eq-Dsigma}
\end{eqnarray}
with $H_\phi$ defined by $\hbar/(4el_\phi^2)$ and $H_W$ given by
$\hbar/(4eW_{eff}^2)$. The effective external magnetic field $B^*$
is defined by:
\begin{equation}
B^*=B \left( 1-\frac{1}{1+W_{eff}^2/3l_B^2} \right) \; ,
\end{equation}
with $l_B=\sqrt{\hbar/eB}$ the magnetic length. The spin-orbit
scattering length $l_{so}$ in the wire can be obtained from the
characteristic spin-orbit field $H_{so}=\hbar/(4el_{so}^2)$.

\begin{figure}[h]
\includegraphics[width=\columnwidth]{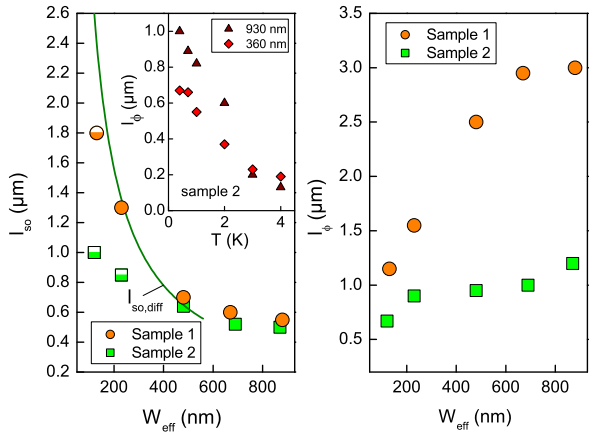}
\caption{(Color online) (a) Spin-orbit scattering length $l_{so}$
determined from the fit of the Kettemann model\cite{Kettemann07}
to the $\sigma(B)-\sigma(0)$ curves at $T=0.4$~K for sample~1
(circles) and sample~2 (squares). The half filled symbols at small
width represent the lower boundary values of $l_{so}$. The inset
shows $l_\phi$ as a function of temperature for the 930~nm and
360~nm wide wires of sample 2. (b) Phase coherence length $l_\phi$
for both samples determined from the fit. \label{Fig5}}
\end{figure}

The Kettemann model was fitted to the experimental curves by
adjusting $H_{so}$ and $H_\phi$. The corresponding values of
$l_{so}$ and $l_\phi$ extracted from the fit are listed in
Table~\ref{Table1} and shown in Fig.~\ref{Fig5}. Even for the widest
wires $l_{so}$ is found to be larger than the value obtained for the
2DEG from the ILP fit. The deviations are probably already due to
confinement effects. In addition, different approximations made in
ILP model\cite{Iordanskii94} for the two-dimensional case and the
Kettemann model\cite{Kettemann07} for wire structures might also
account partially for the deviations.

As can been seen in Fig~\ref{Fig5}, for sample~1 the spin-orbit
scattering length $l_{so}$ monotonously increases with decreasing
$W_{eff}$, while $l_\phi$ decreases. The latter is in accordance
to theoretical predictions.\cite{Altshuler82,Choi87} For the wider
wires with $W=1090$~nm, 880~nm, and 790~nm $l_\phi$ exceeds
$l_{so}$, so that weak antilocalization is expected. In contrast,
for the very narrow wires with $W_{eff}=230$~nm and 130~nm the
values for $l_{so}$ obtained from the fit are close or even exceed
$l_\phi$. In this case the spin-rotation caused by spin-orbit
coupling is not sufficiently strong to affect the interference of
time-reversed paths.\cite{Knap96} As a consequence, the weak
antilocalization effect is suppressed so that weak localization
remains. For the 340-nm-wide wires a satisfactory fit could be
obtained down to a lower boundary value of $l_{so}$, indicated by
the half filled symbol shown in Fig.~\ref{Fig5}(a). In principle,
one could argue, that the appearance of weak localization for the
very narrow wires is solely due to a strongly reduced phase
coherence length, while $l_{so}$ remains at the relatively low
values found for the wider wires. However, in our fits the
suppression of the weak antilocalization effect could not be
explained by simply decreasing $l_\phi$ compared to the values of
the wider wires. A satisfactory fit was only obtained if $l_{so}$
was increased to a larger value compared to the wider wires.

As can be seen in Fig~\ref{Fig5}, for sample~2 the spin-orbit
scattering length $l_{so}$ also increases with decreasing $W_{eff}$,
although with a smaller slope, compared to sample~1. Similarly to
sample~1, $l_\phi $ decreases with decreasing wire width. However,
due to the lower elastic mean free path of sample~2, $l_\phi$ is
considerably smaller for this sample (cf. Fig.~\ref{Fig5}). All
values of $l_{so}$ and $l_\phi$ obtained from the fit are listed in
Table~\ref{Table1}. A comparison of $\sigma(B)-\sigma(0)$ for the
widest wires and for the Hall bar structures reveals, that the weak
antilocalization peak is larger by a factor of two. Thus, although
$l_{el}$ is significantly smaller than $W_{eff}$ this clearly
indicates that the additional carrier confinement already affects
the interference effects.\cite{Beenakker97}

By fitting the Kettemann model to the measurements shown in
Fig.~\ref{Fig4}, $l_\phi$ was determined for the 930~nm and
360~nm~wide wire at different temperatures. For both samples a
fixed value of $l_{so}$, corresponding to the values at 0.4~K,
were assumed. As can be seen in Fig.~\ref{Fig5}(a), inset, for
both samples $l_\phi$ monotonously decreases with temperature, in
accordance with theoretical models.\cite{Altshuler82,Choi87} At a
temperature of 4~K $l_\phi$ is found to be close to $l_{el}$. In
that regime the interference effects are expected to be
suppressed. This is confirmed by the measurements where only a
weak field-dependence of $\sigma(B)-\sigma(0)$ is found.

For both samples we found an increase of $l_{so}$ with decreasing
wire width and even a suppression of weak antilocalization for
narrow wires. This observance is in accordance with weak
antilocalization measurements of quantum wires based on low-band gap
materials, i.e. InGaAs or InAs.\cite{Schaepers06,Wirthmann06}
However, for these types of quantum wells the coupling parameter
$\alpha$ is usually very large. In this case transport takes place
in a different regime where $l_{so} \ll l_{el}$ so that a more
elaborate model had to be applied to extract
$l_{so}$.\cite{Schaepers06} As discussed by
Kettemann,\cite{Kettemann07} the increase of $l_{so}$ can be
attributed solely to a modification of the backscattering amplitude.
In an intuitive picture, the increase of $l_{so}$ in narrow wires
can be explained, by the reduced magnitude of accumulated random
spin phases due to the elongated shape of relevant closed loops.
Here, the spin phase accumulated in forward direction is basically
compensated by the propagation in backwards direction, so that the
spin-related contribution to the interference of electrons
backscattered on time reversed paths tends to diminish. As a result,
only weak localization is observed.\cite{Bournel98,Schaepers06}
Although the spin-orbit coupling strength in our AlGaN/GaN samples
is small compared to heterostructures based on InAs and thus
different models have to be consulted for a detailed description,
the basic mechanism responsible for a suppression of the weak
antilocalization effect is the same for both material systems. In
our case, no decrease of spin-orbit coupling strength, quantified by
$\alpha$, is required to account for the suppression of weak
antilocalization in narrow wires. In fact, an estimation of effect
of the confinement potential on $\alpha$ based on the theory of
Moroz and Barnes\cite{Moroz00} confirmed that for our wire
structures no significant change of $\alpha$ with $W_{eff}$ is
expected. As shown in Fig.~\ref{Fig5} (a), for sample~2 the increase
of $l_{so}$ with decreasing wire width is smaller than for sample~1.
We attribute this to the fact, that for sample 2 the larger extend
of diffusive motion, quantified by the smaller value of $l_{el}$,
partially mask the effect of carrier confinement. Due to the larger
values of $l_{el}$ and $l_\phi$ of sample~1 compared to sample~2,
the shape of the loops responsible for interference effect is
affected more by the confinement of the wire. Thus, the enhancement
of $l_{so}$ is expected to be stronger. Indeed, theoretical
calculations by Pareek and Bruno\cite{Pareek02} showed that for
quasi-onedimensional channels a strong increase of $l_{so}$ can only
be expected if $W_{eff}$ is in the order of $l_{el}$.

For narrow wires with $W_{eff}<l_{so}$ in the diffusive regime
($l_{el}<W_{eff}$) the spin-orbit scattering lengths can be
estimated by:\cite{Kettemann07}
\begin{equation}
l_{so,diff}=\sqrt{12} \frac{l_{so,2D}}{W_{eff}} l_{so,2D} \; .
\label{lsodiff}
\end{equation}
Here, $l_{so,2D}$ is the spin-orbit scattering length of the 2DEG.
The calculated values of $l_{so,diff}$ should only be compared to
the fitted values of $l_{so}$ of sample~2, since only for this
sample $l_{el}<W_{eff}$ is fulfilled. As can be seen in
Fig.~\ref{Fig5} (a), $l_{so}$ calculated from Eq.~(\ref{lsodiff})
fits well to the experimental values corresponding to intermediate
effective wire width of $W_{eff}=480$~nm. However, for smaller
effective wire widths the calculated values of $l_{so,diff}$ are
considerably larger. Probably, spin scattering processes other than
the pure Rashba contribution are responsible for this
discrepancy.\cite{Kettemann07}

An enhanced spin-orbit scattering length is very desirable for spin
electronic devices. Providing that the strength of the spin-orbit
coupling itself remains unchanged, a confinement to a quasi
one-dimensional system would result in a reduced spin randomization.
A reduction of spin randomization is an advantage for the
realization of spin electronic devices, since it would ease the
constraints regarding the size of these type of devices. In this
respect, our finding that $l_{so}$ increases with decreasing wire
width is an important step towards the realization of spin
electronic devices based on AlGaN/GaN heterostructures.

\section{Conclusions}

In conclusion, the magnetotransport of AlGaN/GaN quantum wires had
been investigated. Even for sets of quantum wires with a
geometrical width as low as 340~nm, clear Shubnikov--de Haas
oscillations were observed. Magnetotransport measurements close to
zero magnetic field revealed a suppression of the weak
antilocalization effect for very narrow quantum wires. By
comparing the experimental data with a theoretical model for
one-dimensional structures it was found that the spin-orbit
scattering length is enhanced in narrow wires. The observed
phenomena might have a important implication regarding the
realization of spin electronic devices based on AlGaN/GaN
heterostructures.

The authors are very thankful to S. Kettemann, Hamburg University,
for fruitful discussions and H. Kertz for assistance during low
temperature measurements.


\end{document}